# Preventing detector blinding attack and other strong light attacks on quantum key distribution by use of an explicit random number generator

Mario Stipčević[1][2][3]*

## Introduction

**Availability of random numbers is the key to security of any cryptographic protocol, including quantum key distribution (QKD). A particularly successful "detector blinding" attack [1-2] has been recently demonstrated on various QKD systems, performing for the first time an undetectable and complete recovery of the key. In this paper two original contributions are given to understanding and prevention of this attack. First, it is shown that the detector blinding attack is in fact an attack to receiver's local random number generator (RNG) and that is just a special case of a broader family of RNG attacks which do not necessarily require strong light. Second, based on this insight, an elegant way of preventing a whole family of RNG attacks (including the detector blinding attack) is presented which restores provable security of the system.**

## Results

Although a few different QKD protocols have been broken by the attack, in order to simplify the discussion and without loss of generality, we assume well known BB84 protocol [3] an implementation of which was broken too in [4]. In the BB84 Alice and Bob are linked by one quantum channel and one unsecured but authenticated classical channel. Information from the classical channel can be copied (but not changed) by an eavesdropper (Eve). Over the quantum channel Alice sends to Bob a classical bits (valued 0 or 1) encoded in linearly polarized qubits as follows: 0 is randomly and equiprobably encoded as either $0°$ or $45°$ polarization, while 1 is randomly and equiprobably encoded as either $90°$ or $135°$. Bob detects each qubit in one of the two orthogonal bases ($0°$, $90°$) or ($45°$, $135°$) chosen randomly and with equal probability, using a receiver shown in Fig 1a. Note that Alice needs an equivalent of 2 random bits to generate the qubit state (4 possible states) while Bob needs 1 random bit per received qubit (2 possible bases), which means that each of them must have a random number generator or something equivalent of it.

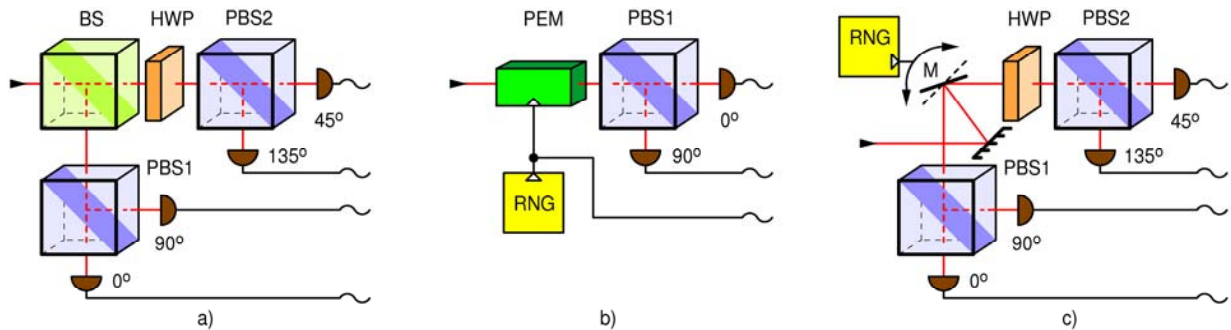

Fig. 1. a) Receiver with a passive random number generator: measurement basis is chosen randomly by means of a first, polarization insensitive beam splitter (BS). Each base consists of a polarizing beam splitter (PBS) and two detectors. b) Active receiver with phase electro-modulator (PEM): measurement basis is determined by the random number generator (RNG) controlling the PEM. c) Exclusive active receiver: receiving basis is

[1]University of California Santa Barbara, Dept. of Electrical and Computer Engineering, Santa Barbara, CA 93106-9530, USA, [2]Duke University Dept. of Physics, Durham, NC 27708 , [3]Ruđer Bošković Institute, Dept. of Experimental Physics, PP 180, HR-10002 Zagreb, Croatia
*e-mail: Mario.Stipcevic@irb.hr

determined by the random number generator (RNG) which flips the mirror (M) between two possible angles each of which reflects *all* light into one and only one measurement basis.

The detector blinding attack works as follows. Eve cuts the quantum channel (e.g. a fiber) and measures Alice's qubit in a randomly chosen base (either ($0^o$, $90^o$) or ($45^o$, $135^o$)) using a receiving station similar to Bob's. We note that in all broken systems Bob's receiving station utilizes a *passive* RNG as shown in Fig. 1a. Eve blinds simultaneously all four detectors by shining strong continuous (CW) circularly polarized light of a carefully tailored intensity. Each detector receives ¼ of the incident power. In that state, detectors are sensitive only to strong pulses of light brighter than some threshold [2], [5], [16]. Eve can therefore make any of the 4 detectors fire at her will by superimposing to the circular CW light a strong pulse of appropriate polarization ($0^o$, $45^o$, $90^o$, $135^o$) that matches her measurement result, so that Bob gets exactly the same result as she did. Next, Eve passively listens to the classical channel between Alice and Bob and does whatever they do in order to arrive to exactly the same "secret" key. (Since blinded detectors do not produce dark counts, a prudent Eve may send to Bob random "noise" pulses thus not taking chances in case that he is monitoring dark count rates of the detectors. Note that by Kerchoff's principle Eve knows the dark count rates of detectors. She could, for example, be the vendor of the QKD station and could have measured dark count rates prior to selling it).

While it is claimed [1], [2], [4] blindability of detectors is a technological weakness that makes this attack viable, our finding is that Eve's success relies solely on the fact that she is able to control Bob's random number generator which is used to choose the detection base. We show this by proving two assertions: 1) If Eve could get an equal control over choice of Bob's detection base in any other way, she would be able to achieve the same result without blinding detectors; 2) If Eve could *not* manipulate Bob's bases then her ability to blind the detectors would not by itself enable her to obtain any information about the key and she would also be discovered. We describe three scenarios which prove the above.

In the first scenario we assume that Eve is able to control Bob's bases by some means other than detector blinding. She chooses her detection basis randomly, measures the qubit sent by Alice, forces Bob's basis to be the same as hers and sends to him a single qubit prepared in the state that she measured. In this way she achieves exactly the same result as she would with the blinding attack but without using the detector blinding. Note that essential part of this attack is her ability to control Bob's choice of bases: not doing anything to the detectors. This proves the first assertion.

In the second scenario Bob uses the setup with an active choice of bases by means of a phase electro modulator (PEM) driven by an explicit electronic random number generator (RNG) as shown in Fig. 1b. The state of the RNG (0 or 1) determines the receiving base. We assume that the RNG is private meaning that it cannot be manipulated nor predicted by Eve. (This setup was used in the original BB84 experimental paper [6]). By mounting the blinding attack described above, Eve is still able to simultaneously blind and keep blinded forever both detectors (and thus both bases) because circularly polarized light distributes evenly among the detectors regardless of the state of the PEM. The difference is that now Eve is able to send her measurement to Bob with only 50% success, namely in those instances where her base coincides with Bob's randomly chosen base by a chance. In the other 50% Bob receives nothing and the bit is lost. By passively listening to the classical communication between Alice and Bob in the next phases of the BB84 protocol, Eve is able to figure out which bits were lost, sift the same bits as Alice and Bob do and recover 100% of the key. While this attack reduces the key rate by a factor of 2 with respect to when Eve is not there, the security situation is not satisfactory because Eve still obtains the key and, in the presence of other losses or a strongly varying loss, she might just get away undetected. Note also that Alice And Bob have no sure way of calibrating the key rate because Eve might decide to be present all the time and generate loss. This system has also been shown vulnerable to the weak pulse (~120 photons) attack with superlinear threshold detectors [7].

The reason for this relative success of Eve even with the active RNG is that this setup is still not satisfying assumptions under which the BB84 protocol has been proven secure (e.g. [8]) namely the light received by Bob's station can may hit *not more than one* base at a time. Note that this does not preclude two or more photons falling onto the *same* base, which is allowed in (proofs of) BB84.

In order to remove this leftover deviation we propose a third scenario involving an improved receiving station, shown in Fig. 1c, that consists of two detection bases and a mirror which directs *all* incoming light to

*one and only one* base. Switching between bases is done under control of an explicit, private RNG. Such a switching mechanism is allowed by laws of quantum mechanics and can be implemented for example by a motorized mirror, a MEMS router [9] or a mechanical switch inside optical fibers [10] all of which can be driven by an electronic quantum random number generator [11-15]. It is the RNG that chooses the receiving bases (not Bob) and Bob is merely informed about the choice so that he can use this information in further phases of the protocol (eg. sifting). Let us now analyze how this setup operates under the detector blinding attack.

Again, Eve shines a strong continuous circular light on Bob's receiver. However now, only one base receives light while the other is in darkness so Eve cannot keep blinded all bases at all times. Whenever Bob's RNG changes its logic state (and thus the receiving base), corresponding detectors that were previously in dark will both simultaneously get hit by a strong light and generate a coincident detection. Not being able to decide which state was sent, Bob will have to announce such events discarded. This will happen in 50% of gates. In the other 50% of gates Eve is able to send her measurement to Bob with probability of ½ (25% of total gates) namely in those instances where her base coincides with Bob's randomly chosen base by a chance and in the rest of 25% gates Bob will receive nothing, resulting in 4-fold drop of effective key rate with respect to no eavesdropper. Even though this is better than in the previous scenario it is still not satisfactory because Eve again has a complete knowledge of the key. However, there is another lever arm available to Alice and Bob. In order to be sensitive to BER, the receiving station should keep the per-gate noise detection at the level of ~1% thus limiting the accidental coincident detection probability $p_c$ during single photon communication to $\sim 10^{-4}$. It is easy for Bob to detect the huge rise in coincidence probability to ~ ½ when under the "strong light" attack of some kind. Note that the per-gate probability of coincident detection will persist close to ½ even if blinding power drops down to a few photons per gate and per detector because the detection probability $p_d$ of a detector of efficiency $\epsilon$ hit by $N$ photons is:

$$p_d = 1 - (1 - \epsilon)^N \qquad (1)$$

or closer to unity in case of superlinear detectors [7]. Assuming realistic value $\epsilon = 0.25$ and total incoming light power of only $2N$ = 20 photons (per gate), the coincidence probability during communication given by:

$$p_c' = \frac{1}{2} p_d^2 \qquad (2)$$

is about 0.45. This is still orders of magnitude higher than expected while the incident power is much smaller than required for an actual blinding attack. Unlike the undisturbed key rate, Bob can confidently calibrate the coincidence rate of the undisturbed receiving station by simply disconnecting it from quantum channel and measuring the coincidence probability $p_{c0}$ which is slightly smaller or equal than $p_c$.

In order to overcame this defense, Eve might try to attack only during a portion of gates and thus obtain smaller portion of the key but with less disturbance. However, on average, with every bit acquired Eve generates two coincidences. By measuring the coincidence probability $p_c'$ in communication via quantum channel and comparing it to the undisturbed probability $p_{c0}$ and assuming that the number of bits known to Eve is ½ of the number of extra coincidences, Bob can calculate the upper bound on number of bits leaked to Eve and account for it by stronger privacy amplification.

In any scenario we see that Eve has knowledge on only those bits for which she was able to control (and thus know) the state of the local RNG which proves the assertion 2.

**Conclusion**

Although passive (Fig. 1a) and active (Fig. 1c) receiving schemes are seemingly functionally equivalent, the latter is secure against RNG attacks. Namely, even though an information-theoretically perfect random number generator, required for choosing of the receiving bases, can be realized as a part of the passive receiving station (namely the photon source, the non-polarizing beam splitter and detectors [12]), a subtle difference is that while and implicit RNG is a part of the optical system of the receiving station and is thus accessible by an attacker, an explicit RNG internally collapses the wave function and outputs only classical information. Such a RNG does not accept any information from communication channels and consequently cannot be manipulated by Eve. Research of countermeasures to blinding attacks has so far concentrated on preventing the specific features of blindability itself [1],[2],[17-21] while having no bearing on other potential weaknesses. We believe that our approach that is non-specific to the blinding attack and yet restores the provable security should have a higher resilience to wider range of threats that may be conceived in the future.